\documentclass[prl,twocolumn,showpacs,floatfix,preprintnumbers,amsmath,
amssymb]{revtex4}

\usepackage{graphicx}
\usepackage{amsmath}
\usepackage{natbib}
\newcommand{\ve}{\varepsilon}

\begin{document}
\title{Coherent control of interacting electrons in quantum dots by 
traveling the spectrum}
\author{G.\ E.\ Murgida}
\author{D.\ A.\ Wisniacki}
\author{P.\ I.\ Tamborenea}
\affiliation{Departamento de F\'{\i}sica ``J.\ J.\ Giambiagi",
Universidad de Buenos Aires, Ciudad Universitaria, Pab.\ I,
C1428EHA Buenos Aires, Argentina}
\date{\today}

\begin{abstract}
Quantum control of the wave function of two interacting electrons confined 
in quasi-one-dimensional
double-well semiconductor structures is demonstrated. 
The control strategies are based on the knowledge of the energy spectrum 
as a function of an external uniform electric field. 
When two low-lying levels have avoided crossings our 
system behaves dynamically to a large extent as a two-level system. 
This characteristic is exploited to implement coherent control strategies
based on slow (adiabatic passage) and rapid (diabatic Landau-Zener
transition) changes of the external field. 
We apply this method to reach desired target states that lie far in the 
spectrum from the initial state.
\end{abstract}


\maketitle

The control of quantum systems is a fundamental challenge in physical
chemistry, nanoscience, and quantum information 
processing \cite{wal-rab,ric-zha}.
Quantum control is the manipulation of the temporal evolution of a system
in order to obtain a desired target state or value of a certain physical
observable.
From the experimental point of view, the techniques of quantum control 
are highly developed in the area of magnetic resonance, and more
recently great progress has been made in quantum chemistry thanks to 
the development of ultrafast laser pulses \cite{zew}.

Coherent control in semiconductor quantum dots has become an active
field of research in the last 15 years.
Early works on electron localization in double well systems spurred
intense theoretical activity.
In a seminal paper, Grossmann {\it et al.} \cite{gro-dit-jun-han} showed 
that, by applying an appropriate AC electric field, the tunneling of 
the electron between the wells could be coherently destroyed, thereby
maintaining an existing localization in one of the wells.
Shortly after, Bavli and Metiu \cite{bav-met} found ways to, starting 
from the delocalized ground state, localize the electron wave function 
and then to preserve the localization with a precisely taylored 
time-dependent electric field. 
A large body of literature followed these pioneering works.
A decade later, the first steps in the theoretical exploration of 
localization and control of two interacting electrons in quantum dots 
were made \cite{tam-met-99,zha-zha-00,tam-met-01}.
Whereas Zhang and Zhao studied a model two-level system, Tamborenea and 
Metiu studied a more realistic multi-level system inspired by 
quasi-one-dimensional semiconductor nanorods.
The study of two-electron localization and control in double dots has 
remained active ever since \cite{zha-zha-01,pas,wan-dua-zha,pas-ter,cre-pla}. 

In this Letter we propose an efficient method to control 
the wave function of two interacting
electrons confined in quasi-one-dimensional nanorods \cite{mas-bie-mar,
tam-met-01}. 
The control method is based on the knowledge of the energy spectrum as 
a function of an external uniform electric field. 
The method requires that the system behaves locally---near avoided level
crossings---as the Landau-Zener (LZ) two-level model \cite{zen}. 
This fact is exploited to navigate the spectrum using slow (adiabatic) and
rapid (diabatic) changes of the external control parameter.
Although this characteristic may seem rather restrictive, it is, in fact,
a general property of systems with interaction between its energy levels.
The level repulsion must not be too strong, though, so that the spacing
at the avoided crossings remains smaller than the mean level spacing.

Let us consider a quasi-one-dimensional double quantum dot with two 
interacting electrons in it in the presence of a spatially uniform electric 
field \cite{tam-met-01}. 
Denoting by $z$ the longitudinal coordinate,
the Hamiltonian of the two electrons reads
\begin{eqnarray}
H&\equiv&-\frac{\hbar^2}{2m}
      \left(\frac{\partial^2}{\partial z_1^2} + 
            \frac{\partial^2}{\partial z_2^2}
      \right) + V(z_1) + V(z_2) \nonumber
 \\
&&+ V_C(|z_1-z_2|) - e(z_1 + z_2)E(t),
\label{eq:hamiltonian}
\end{eqnarray}
where $m$ is the effective electron mass in the semiconductor material, 
$V_C$ is the Coulomb interaction between 
the electrons, $V$ is the confining potential, and $E(t)$ is an external 
time-dependent electric field.
The confining potential is a double quantum well with well width of 28~nm,
interwell barrier of 4~nm, and 220~meV deep (a typical depth for a GaAs-AlGaAs
quantum well).   
We mention that the control results that we report here are robust 
with respect to the fine tuning of the parameters of the structure.
We normally assume that the initial state is the ground state, which is 
a singlet.
Since the Hamiltonian is spin independent, the total spin is conserved 
and the spatial wave function remains symmetric under particle exchange
at all times.

\begin{figure}[hbp]
\includegraphics[width=7.cm,angle=-90]{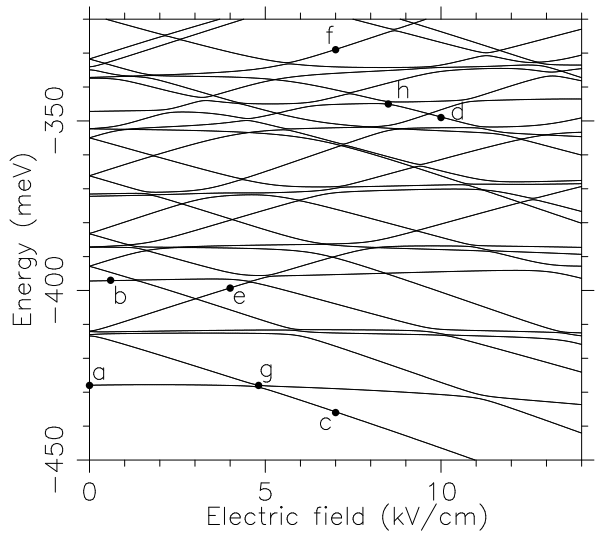}
\caption{Energy spectrum of our system as a function of the
external uniform electric field.}
\label{fig:espectro}
\end{figure}

We first consider the case of constant electric field, to be considered at
this point as a parameter in the Hamiltonian.
In Fig.\ \ref{fig:espectro} we show the spectrum of eigenenergies of the
Hamiltonian of Eq.\ (\ref{eq:hamiltonian}) versus electric field.
The energies $\epsilon_i(E)$ and  eigenstates $\phi_i(E,z_1,z_2)$
are obtained by numerical diagonalization.
We have used as basis set the symmetric combinations of 
the twelve lowest single-particle eigenfunctions. 
Thus, our basis of the two-particle Hilbert space has 12*(12+1)/2=78 states.
We can see in Fig.\ \ref{fig:espectro} that the spectrum is composed of 
nearly straight lines \cite{footnote}. 
A closer look reveals that all level crossings are avoided, giving rise 
to adiabatic curves that never cross each other. 
The fact that all crossings are avoided ones is due to the electron-electron 
interaction, which couples the energy levels of the non-interacting
system \cite{fen-san-tam}.

\begin{figure}[hbp]
\includegraphics[width=4.3cm,angle=-90]{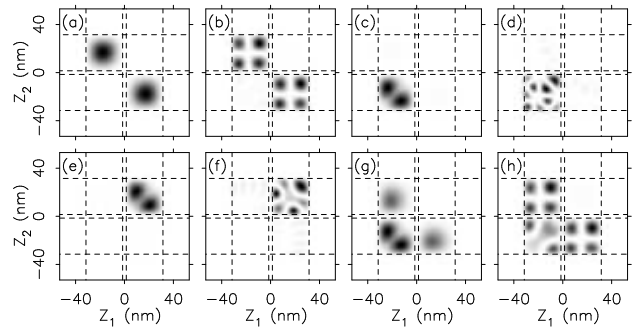}
\caption{
Spatial wave functions $\phi_i(E,z_1,z_2)$ corresponding to labels 
(a) to (h) of Fig.\ \ref{fig:espectro}.
States (a) and (b) have one electron in each well, wave functions 
(c) and (d) are localized in the left well, and wave functions (e) and (f)
have both electrons in the right well. 
States (a) to (f) are far from avoided crossings and therefore have 
well-defined localization properties. 
This is not the case for eigenstates (g) and (h), which are at avoided 
crossings. 
}
\label{fig:wavefunctions}
\end{figure}

The straight lines of the spectrum are distributed in three clearly 
distinguishable groups: 
those with negative, zero and positive slope. 
In each group the slopes are very similar, and far from the avoided crossings, 
the wave functions have a distinct kind of localization, as follows
(see Figs.\ \ref{fig:espectro} and \ref{fig:wavefunctions}(a-f)): 

\noindent
i) for zero slope the electrons are delocalized, that is, each 
electron is in a different dot (Fig.\ \ref{fig:wavefunctions}(a,b)),\\ 
ii) for negative slopes both electrons are in the left dot 
(Fig.\ \ref{fig:wavefunctions}(c,d)), and \\
iii) for positive slopes both electrons are in the right dot 
(Fig.\  \ref{fig:wavefunctions}(e,f)).

\noindent
Along a given straight line the eigenstates do not change much, thus each line
has associated a characteristic shape of the wave function.
Near an avoided crossing, states with different types of localization mix.
These mixed states do not have a well-defined localization type.
As an example, in Fig.\ \ref{fig:wavefunctions}(g) and (h) we show the 
eigenstates 
at the avoided crossing between the first two levels at $E=4.81$ kV/cm and 
between the levels 21 and 22 at $E=8.48$~kV/cm, respectively. 


Our goal is to find a method to control the wave function.
The previous discussion about the spectrum and the characteristics 
of the wave functions at and far from the avoided crossings 
suggests a possible control strategy.
For example, starting at state $(a)$ of Fig.\ \ref{fig:espectro} and 
varying slowly (adiabatically) the electric field we reach state $(c)$, 
which has a different type of localization (see Fig. \ref{fig:wavefunctions}).
On the other hand, if we vary the parameter $E$ quickly the final state 
will have the same localization as the initial one (see below, 
Fig.\ \ref{fig:nav1}).
These types of transitions will be the building blocks of our control strategy.
The exact meaning of slow and fast in this context is given by the
LZ model \cite{zen}.

The LZ theory treats a rather simple and generic situation of two-level
avoided crossing. The LZ Hamiltonian in the 
diabatic basis $\{|+\rangle,|-\rangle\}$ is
\begin{equation}
 H=\left[\begin{array}{cc}
        \varepsilon_+(\lambda) & \varepsilon \\
        \varepsilon            & \varepsilon_-(\lambda)
     \end{array}\right]
 \label{eq:hamiltonianLZ}
\end{equation}
where $\varepsilon$ is a constant while the diagonal  $\ve_{\pm}(\lambda) = E_0 + \alpha_{\pm} \lambda$ are linear 
functions of the parameter $\lambda$.
It is assumed that the parameter $\lambda$ varies linearly 
with time. 
It can be shown that the probability of remaining in state $|+\rangle$ 
is given by
\begin{equation}
P_+ = \exp{\left[
           \frac{-2\pi \ve^2}
                {\hbar (\alpha_+ - \alpha_-) \dot{\lambda}
                }
          \right]}, 
\label{eq:probaLZ}
\end{equation}
where $\dot{\lambda}$ is the rate of variation of the parameter.
Conversely, $P_- = 1 - P_+$.
By selecting the rate of change of the parameter $\lambda$ 
we can control the final state of the system. For slow variations of   
$\lambda$ ($\dot{\lambda} \ll \frac{\ve^2}{\hbar}$),
the system follows the adiabatic curve passing from the initial 
diabatic state to the other one. 
In the opposite case, when $\dot{\lambda} \gg \frac{\ve^2}{\hbar}$,
the evolution takes place on the diabatic curve and the state remains as 
the initial one. 
It can be shown that near an avoided crossing our system can be treated as a 
two-level LZ model \cite{mur-wis-tam}.
We compared the evolution of the two-level LZ Hamiltonian  
(Eq.\ (\ref{eq:hamiltonianLZ})) to that of the complete Hamiltonian
(Eq.\ (\ref{eq:hamiltonian})) for many rates of variation of the parameter
(electric field) and we found very close agreement. 
The numerical solution of the time-dependent Schr\"odinger equation was 
obtained using the usual fourth-order Runge-Kutta method \cite{tam-met-99,tam-met-01}.
We use a time step of $0.1 \, \mbox{fs}$, which ensures that the precision 
of all the reported probabilities and overlaps is better than $0.4 \%$.

The adiabatic transition from state (a) to (c) (see Figs.\ \ref{fig:espectro} 
and \ref{fig:wavefunctions} (a,c)) is an extremely simple solution to the problem of 
localization in a realistic double quantum dot \cite{tam-met-01,zha-zha-00,
zha-zha-01}. 
In fact, more complex and interesting control problems can be tackled. 
Namely, by suitable combinations of rapid (diabatic) and slow (adiabatic) 
changes of the electric field it is possible to travel over the spectrum 
connecting distant pairs of states. 
That is, we can not only control the distribution of electrons in the 
dots (localization and delocalization), but we are also able to choose finer 
details of the final state, like, for example, its nodal domains.

\begin{figure*}
 \includegraphics[width=14.0cm]{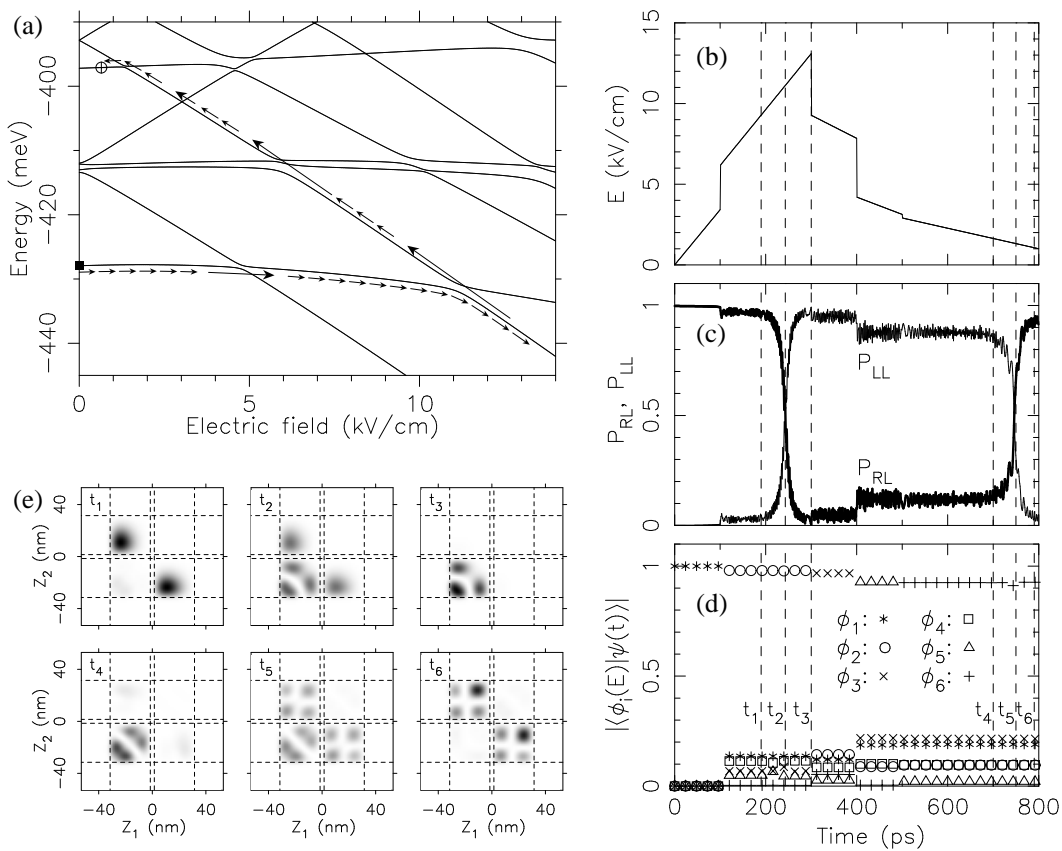}
 \caption{(a) Schematic diagram of the intended path followed by the state.
Short (long) arrows indicate slow (fast) variations of the electric field.
The initial state is marked as $\blacksquare$ and the target state as $\oplus$.
(b) External electric field (the control parameter) as a function of time.
(c) Probabilities that the two electrons are in the same (left) well 
($P_{LL}$) and in different wells ($P_{RL}$).
(d) Absolute value of the overlap of the evolving state with the energy 
eigenstate $\phi_i(E)$.
(e) Spatial wavefunctions at various times indicated in (d).}
 \label{fig:nav1}
\end{figure*}

We now show a complex navigation through the spectrum in which
the system is taken from the ground state to a specific high-energy 
state.
In this case, the target is state (b) of Fig.\ \ref{fig:wavefunctions}
(see also Fig.\ \ref{fig:espectro}), which shares the localization type with 
the ground state (they are both delocalized), but has a more complex nodal 
structure. 
This process is depicted in Fig.\ \ref{fig:nav1}.
We want to reach the excited state by means of diabatic and adiabatic
transitions.
The intended navigation path is displayed with arrows in the
spectrum of Fig.\ \ref{fig:nav1}(a).
The small arrows indicate slow variations of the control electric field
whose objective is to follow the adiabatic states (i.e.\ the eigenfunctions of
the Hamiltonian at the successive values of the electric field). 
The long arrows denote diabatic transitions at the avoided crossings.
Here we use instantaneous jumps of the electric field, but we
have checked that risetimes of the order of 0.1~ps do not change 
significantly our results.
The time dependence of the electric field is shown in Fig.\ 
\ref{fig:nav1}(b).
In order to know in detail how the evolution of the wave function 
proceeds, we show in Fig.\ \ref{fig:nav1}(c,d) different
aspects of the time-dependent wave function.
In Fig.\ \ref{fig:nav1}(c) we compute the time-dependent localization
probabilities, $P_{LL}$, that the two electrons are in the left well,
and $P_{RL}$, that they are in different wells.
Fig.\ \ref{fig:nav1}(d) gives the absolute value of the overlap of 
the evolving state with the energy eigenstate $\phi_i(E)$, and
Fig.\ \ref{fig:nav1}(e) displays the spatial wavefunction at the 
various times indicated in (d).

Let us discuss in detail the behavior of the evolving wave function.
We start from the ground state with zero electric field and move 
adiabatically up to a field of 3.5 kV/cm, as shown 
in Fig.\ \ref{fig:nav1}(a).
This process takes 100 ps (see Fig.\ \ref{fig:nav1}(b)).
Note in Fig.\ \ref{fig:nav1}(d) that the overlap with the 
adiabatic eigenstate $\phi_1$ is approximately 1.
After that the field is quickly increased (the first jump of the electric
field in Fig.\ \ref{fig:nav1}(b), at $t = 100$ ps) so that the
system goes through the first avoided crossing. 
At this avoided crossing the overlap between the evolving state and the
adiabatic eigenstates begins to decay, from almost 1 to close to 0.97.
For example, at time $t_1 = 190$ ps the wave function still resembles 
the ground state (see Fig.\ \ref{fig:nav1}(e)) but displays a degree 
of mixture with localized states, as evidenced by the finite probability 
in the lower-left quadrant.
Afterwards, we move slowly until $t_3 = 300$ ps passing adiabatically
the second avoided crossing at $E = 11.2$ kV/cm.
The mixed nature of the wave function at the avoided crossing 
(at $t_2 = 243$ ps) can be seen in Fig.\ \ref{fig:nav1}(e).
After the crossing, at $t_3$, the wave function is highly localized on the 
left dot (see Fig.\ \ref{fig:nav1}(e)).
At this point, we still have ahead of us a long way to the 
desired final state, most of it along a spectral "line" of negative slope. 
Although it would be tempting to make a sudden change of the control 
parameter to its final value, that strategy is not satisfactory.
It is best to proceed slowly far from the avoided crossings and rapidly
around them.
This procedure has the advantage of allowing our state to adjust to the
gradual changes of the adiabatic states in the regions between crossings.
These gradual changes do not entail changes in the type of localization,
but rather displacements of the probability density within the quadrants
which are already populated.
In the remainder of the path we cross three avoided crossings diabatically
and the last one slowly.
The latter, being very narrow, turned out to be the hardest one to 
pass adiabatically.
This passage took 300 ps, as seen in Fig.\ \ref{fig:nav1}(b).
In this avoided crossing the wave function goes from a strong localization
in the left dot (wave function at $t_4$ of Fig.\ \ref{fig:nav1}(e))
to a clear delocalization in the final state.
We remark that the overlap of the final state with the desired state, at
$t_6 = 800 \, \mbox{ps}$, is 0.93 (see Fig.\ \ref{fig:nav1}(d))

Before concluding, it is worth comparing the present control method 
with others found in the literature. 
In Ref.\ \cite{tam-met-01}, a different localization scheme was 
proposed for the system studied here.
In that work the electron localization is obtained, starting from the 
ground state at zero electric field, with a piecewise constant electric
field.
The time scales involved, of a few picoseconds, which are experimentally 
accesible, are the same in the two approaches.
However, the present method has the advantage of being more robust since 
it does not require the fine tuning of the field jumps and, furthermore, 
it gives higher probabilities of reaching the target state.
Moreover, in Ref.\ \cite{tam-met-01} the controlled state is a superposition
of only the two lowest-energy eigenstates, while in our proposed method 
a much more general navigation of the spectrum is naturally possible,
allowing to connect very distant states.
More recently, an efficient method of control was proposed and applied to
the creation of entangled states in spin systems \cite{una-vit-ber}. 
The method is based on the navigation of the energy spectrum, expressed
as a function of an external control parameter, which is varied always
adiabatically.
A key element of the controlled system of that study is that the interaction,
which controls the nature of the level crossings, can be switched off at will.
This additional control tool is clearly not generic. 
Here we propose a control method based on similar general ideas but 
which is more general in the sense that the interaction among levels 
does not need to be controlled. 
Finally, we mention that nonadiabatic LZ transitions have been applied 
to control single-electron transport \cite{mak}, and adiabatic passage
has been used to control localization and suppression of tunneling
\cite{san-gue-amn-jau}.
 
In summary, we have proposed an efficient method to control the wave function 
of a two-electron quantum dot system.
This method is based in slow and fast variations of a control parameter
which in our study is given by an external electric field that couples
to the electrons through the dipole Hamiltonian.
The success of this method relies on the condition that the
interaction between neighboring levels be well described by the two-level 
LZ model.
Although this may seem a severe limitation to the applicability of the 
method, this condition is quite generally satisfied. 
For example, it has been recently shown in the paradigmatic stadium billiard 
that the transitions between neighboring levels are of the LZ
type when the billiard boundary is deformed \cite{san-ver-wis}. 
We have shown that with this method an effective control of the wave function
can be achieved, even for rather ambitious control goals.
While the technology to produce the required ramped electric fields,
controllable in the picosecond time scale with enough flexibility, may
not be currently available, we expect that our proposal will further
motivate their experimental development.

The authors acknowledge support from CONICET (PIP-6137,PIP-5851) and 
UBACyT (X248, X179). DAW and PIT are researchers of CONICET.



\begin{thebibliography}{99}
\bibitem{footnote} The slopes are approximately given by the mean value
of $z_1+z_2$, or, in other words, the dipolar moment of the wave functions.
\end{thebibliography}
\end{document}